\documentclass[sigconf]{acmart}
\usepackage{tikz}\usepackage{tcolorbox}
\usepackage{booktabs}
\usepackage{graphicx}
\usepackage[skip=1pt]{caption}
\usepackage{setspace}
\usepackage{tabularx}
\usepackage{booktabs}
\usepackage{array}
\usepackage{enumitem}

\usetikzlibrary{positioning,arrows.meta}
\AtBeginDocument{%
  }

\setcopyright{acmlicensed}
\copyrightyear{2018}
\acmYear{2018}
\acmDOI{XXXXXXX.XXXXXXX}
\acmConference[Conference acronym 'XX]{Make sure to enter the correct
  conference title from your rights confirmation email}{June 03--05,
  2018}{Woodstock, NY}
\acmISBN{978-1-4503-XXXX-X/2018/06}
\usepackage{xspace}
\usepackage{tcolorbox}

\begin{document}

\title[Repository-Aware Metamorphic Relation Generation for Augmented Reality Applications using LLMs]{Repository-Aware Metamorphic Relation Generation for Augmented Reality Applications using Large Language Models}

\author{Dibyendu Brinto Bose}
\orcid{0000-0002-1603-3574}
\affiliation{%
  \institution{Virginia Tech}
  \city{Blacksburg, VA}
  \country{USA}}
\email{brintodibyendu@vt.edu}

\author{Jiawei Qin}
\orcid{0009-0008-6425-2019}
\affiliation{%
  \institution{Virginia Tech}
  \city{Blacksburg, VA}
  \country{USA}}
\email{jiaweiq@vt.edu}

\author{Chris Brown}
\orcid{0000-0002-6036-4733}
\affiliation{%
  \institution{Virginia Tech}
  \city{Blacksburg, VA}
  \country{USA}
}
\email{dcbrown@vt.edu}






\renewcommand{\shortauthors}{Bose et al.}

\begin{abstract}
Metamorphic Testing (MT) provides a promising approach for testing software without defined test oracles by specifying expected relations between inputs and outputs, instead of relying on exact outputs.  For example, testing Augmented Reality (AR) applications is challenging due to dynamic interactions between virtual content, physical environments, and code, which make traditional test oracles difficult to define. However, formulating metamorphic relations (MRs) is time-consuming and burdensome. We introduce a context-aware pipeline that generates and refines MRs using repository-level context and reasoning orchestration, evaluated on a dataset of 142 mobile AR system repositories. Across three context configurations generating 14{,}916 candidate MRs, hierarchical context yielded the broadest coverage (7{,}004 MRs across 142 repositories and 5{,}167 class--method pairs) and lower redundancy. An agentic deliberation process then reconciled conflicting candidates—observed in 79.0\% of cases—reducing duplication and selecting context-aware relations in 88.2\% of outcomes. A manual oracle study shows refined relations ($n = 141$) are both logically valid and sufficiently concrete to be directly translated into test assertions, and a preliminary case study reveals converting generated MRs ($n = 5$) into executable tests can detect non-equivalent mutations in real-world code. Overall, our results show that combining repository-aware MR generation with reasoning-based refinement enables scalable construction of reliable, domain-relevant test oracles. 

\end{abstract}



\maketitle

\section{Introduction}\label{sec:introduction}







Metamorphic Testing (MT) provides a practical solution for scenarios where defining precise test oracles is difficult. Rather than verifying single outputs, MT defines \emph{metamorphic relations} (MRs) that describe how program outputs should change when inputs are transformed~\cite{mr_base, altamimi2023metamorphic}. Many complex software systems exhibit such relations. For example, Augmented Reality (AR) applications increasingly integrate interactions between digital content, physical environments, and code to provide users with immersive experiences across critical domains~\cite{badimpchi, Kraus_CHI, suddenchi}. The output of AR programs often depends on dynamic environmental conditions, spatial information, camera input, and device motion~\cite{badimpchi, suddenchi}, making it difficult to determine expected outputs for test cases~\cite{bose_ar_mt}. Yet, such properties can be encoded as relations to validate AR application behavior~\cite{bose_ar_mt}. For example, when the camera moves closer to a virtual object in an AR scene, the rendered object should appear proportionally larger, and when the camera moves away it should appear smaller. 

However, identifying meaningful MRs is challenging across software domains. For instance, MR generation often requires substantial domain expertise~\cite{mr_deeplearning}. Developers must reason about how input transformations affect system behavior and how these relations can be operationalized into executable tests~\cite{mr_stateofart}. In large software repositories containing hundreds of classes and thousands of methods, manually identifying such relations becomes difficult and time-consuming~\cite{li2025mrg}.

Recent advances in large language models (LLMs) have created new opportunities to automate test specification generation~\cite{metal_mr}. Prior studies show that LLMs can assist in generating code tests and reasoning about program behavior~\cite{celik2025review, li2025mrg}. Nevertheless, existing approaches typically prompt LLMs using isolated functions or short code snippets, limiting the model's ability to understand the broader architectural context in which a method operates~\cite{LLM_test, llm_test2}. Another challenge is that LLMs lack program context for testing tasks~\cite{rao_cat-lm_2023}, causing them to generate incorrect or irrelevant test cases that fail to verify program behavior~\cite{mathews2024design}.

To address these challenges, recent work suggests \textit{repository-level code understanding} aids in effective reasoning about software systems, requiring structured contextual information beyond individual files~\cite{repomain, hoang2025codewiki, celik2025review}. For example, Oskooei et al. propose representing repositories through hierarchical summaries to enable LLMs to reason about large codebases while mitigating context window limitations~\cite{oskooei2025repository}. However, increasing context introduces a new challenge: different configurations often produce competing or overlapping MRs for the same target method, varying in specificity, domain relevance, and sometimes even contradicting each other~\cite{mr_coverage_new, mr_selection_good}. Therefore, beyond generating MRs, it is necessary to systematically reconcile these differences to obtain reliable relations. Prior work shows \textit{debate-based refinement} reduces contradictions and redundancies in LLMs for various software engineering tasks, such as code generation~\cite{islam2025codesim} and issue resolution~\cite{li2025swe}.

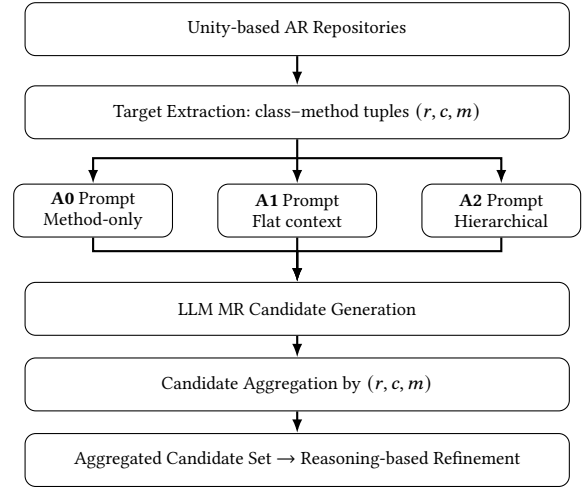
\begin{figure}[t]
\centering
\footnotesize
\begin{tikzpicture}[
    block/.style={draw, rounded corners, align=center,
                  minimum width=7.2cm, minimum height=0.7cm,
                  font=\footnotesize},
    sblock/.style={draw, rounded corners, align=center,
                   minimum width=2.1cm, minimum height=0.7cm,
                   font=\footnotesize},
    line/.style={-{Latex[length=1.8mm]}, thick}
]

\node[block] (repos) at (0,0)
    {Unity-based AR Repositories};

\node[block] (extract) at (0,-1.1)
    {Target Extraction: class--method tuples $(r,c,m)$};

\node[sblock] (a0) at (-2.7,-2.4) {\textbf{A0} Prompt\\Method-only};
\node[sblock] (a1) at (0,-2.4)    {\textbf{A1} Prompt\\Flat context};
\node[sblock] (a2) at (2.7,-2.4)  {\textbf{A2} Prompt\\Hierarchical};

\node[block] (gen) at (0,-3.7)
    {LLM MR Candidate Generation};

\node[block] (agg) at (0,-4.7)
    {Candidate Aggregation by $(r,c,m)$};

\node[block] (out) at (0,-5.7)
    {Aggregated Candidate Set $\rightarrow$ Reasoning-based Refinement};

\draw[line] (repos) -- (extract);
\draw[line] (extract.south) -- ++(0,-0.25) coordinate (branch)
    -- (branch -| a1.north) -- (a1.north);
\draw[line] (branch) -- ++(0,0) -| (a0.north);
\draw[line] (branch) -- ++(0,0) -| (a2.north);
\draw[line] (a0.south) |- ++(0,-0.18) -| (gen.north);
\draw[line] (a1.south) -- (gen.north);
\draw[line] (a2.south) |- ++(0,-0.18) -| (gen.north);
\draw[line] (gen) -- (agg);
\draw[line] (agg) -- (out);

\end{tikzpicture}
\caption{MR generation pipeline. For each target repository, class, and method tuple $(r,c,m)$, prompts
are constructed under A0 (method-only), A1 (flat context), and A2
(hierarchical context), passed to a shared LLM, and aggregated for
downstream refinement.}
\label{fig:mr_generation_pipeline}
\end{figure}

Motivated by these insights, we explore whether structured repository context can improve the generation of MRs. Critically, existing work on debate-based LLM refinement has focused on \emph{detecting} whether a candidate relation holds for a given code snippet~\cite{BOSE_LLM_IDENTIFICATION}, a classification problem where ground-truth relations are assumed to exist. The present work addresses an orthogonal problem: \emph{generating} MRs from scratch using repository-level context, where no ground-truth relation exists, and the challenge is both discovery and quality. We organize context into three \sloppy \mbox{levels—repository-level} metadata, class-level semantics, and method-level signatures—to expose both architectural intent and localized implementation details relevant for behavioral reasoning. We also introduce a reasoning-orchestrated deliberation mechanism that identifies contradictions and redundancies among candidates and refines them into more coherent MRs. Figure~\ref{fig:mr_generation_pipeline} provides an overview of the complete pipeline. Our study is guided by the following research questions:


\begin{itemize}[topsep=0pt]
\item \textbf{RQ1:} How does repository-level contextual granularity affect the structural validity and coverage of LLM-generated MRs?

\item \textbf{RQ2:} To what extent can structured deliberation reconcile competing MR candidates generated from different context configurations?

\item \textbf{RQ3:} Can debate-refined MRs serve as executable 
test oracles for AR applications, and how does deliberation 
strategy affect their quality?
\end{itemize}
We evaluate our approach on a sample of real-world AR systems due to the complexity and difficulty of defining test oracles for AR applications~\cite{badimpchi}. Our results show that hierarchical repository context improves coverage, specificity, and diversity of generated MRs while maintaining high structural validity. Importantly, the relationship between context volume and quality is not monotonic: flat retrieval (A1) degrades source traceability relative to 
both the method-only baseline and hierarchical context, showing that context \emph{organization} matters as much as context \emph{volume}. Furthermore, the proposed deliberation mechanism effectively resolves conflicts among candidates in 90\% of cases, reduces exact duplicate rates from 13.3\% to 1.3\%, and 
produces more precise and context-aware relations — with merged outputs outperforming single-best selections on every quality dimension. Finally, manual inspection shows that debate-refined MRs achieve high validity (1.80/2) and testability (1.82/2), with 61.7\% rated high-quality across all dimensions. 
Through this study, we aim to better understand how structured repository-level context influences LLM-based test specification generation and whether reasoning orchestration mechanisms can improve the reliability of generated relations. The main contributions of this paper are:

\begin{itemize}[topsep=0pt]

\item An empirical study demonstrating how repository-level contextual granularity affects the coverage, validity, and diversity of generated MRs for AR applications, revealing that context organization matters as much as context volume.


\item A reasoning-orchestrated deliberation framework that resolves contradictions and redundancies among candidate MRs and produces refined, context-aware relations.

\item An oracle study evaluating the validity, testability, and AR-specific relevance of curated MRs, providing insights into their usefulness as test specifications.

\end{itemize}

\section{Background}

\subsection{Challenges in Identifying MRs}

Identifying MRs remains a challenging task. Constructing meaningful MRs typically requires deep understanding of both the application domain and the implementation of the software system~\cite{mr_base}. Developers must determine which input transformations are meaningful, how outputs should change under those transformations, and how such relations can be operationalized into executable tests~\cite{chen2021evaluating, austin2021program}. The difficulty increases in large software repositories. Modern systems often contain hundreds of classes and thousands of methods, making it difficult for developers to systematically identify behavioral relations across different parts of the system~\cite{mr_largerepo1, mr_largerepo2}. Moreover, many relations depend on architectural context, such as interactions between classes or how specific methods manipulate spatial data within the AR pipeline~\cite{mr_deeplearning}.

Prior approaches leverage search-based~\cite{zhang2014search}, pattern-based~\cite{segura2018metamorphic}, and mutation-based~\cite{shan2009generating,sun2024identifying} approaches. Recent advances in LLMs have created new opportunities to assist developers in generating testing artifacts such as test cases and specifications~\cite{LLM_test, tufano2022using}. Prior work has investigated using LLMs to determine whether a given code snippet conforms to predefined MRs~\cite{BOSE_LLM_IDENTIFICATION}, including deducing MR input transformations for MT~\cite{xu2024mr}. However, these approaches assume that MRs are already known and therefore do not address the problem of discovering new relations directly from repository-level program context.

In contrast, our work addresses the previously underexplored problem of \emph{MR discovery from repository-level program context}. Unlike prior approaches that operate on isolated snippets or predefined relations, we leverage structured repository context to enable LLMs to reason about behavioral properties across classes, methods, and dependencies. Furthermore, we introduce a reasoning-orchestrated deliberation mechanism to reconcile competing candidate MRs, a step that is absent in existing MR generation pipelines. This combination of \emph{context-aware MR generation} and \emph{deliberative refinement} allows us to produce more specific, domain-relevant, and operationalizable relations for AR systems.

\subsection{Testing Challenges in AR Systems}

Our work explores repository-level context for MR generation in the context of AR programs. AR applications operate in highly dynamic environments where program behavior depends not only on internal logic but also on external sensor inputs such as camera frames, spatial tracking, and device orientation~\cite{Kraus_CHI, suddenchi}. These characteristics introduce testing challenges that are less common in traditional software systems. For example, the output of an AR application may depend on environmental conditions such as lighting, surface detection, and camera movement, making it difficult to define precise expected outputs for test cases~\cite{badimpchi}. When tests can be implemented, prior work shows testing code for mixed reality applications suffers from substantial test smells~\cite{gurramkonda2025vrtestsniffer}.

A persistent challenge for AR systems is the \emph{test oracle problem}, where determining the correct output for a given input is challenging or infeasible~\cite{daart}. While exact outputs are often unknown, developers may still reason about expected \emph{relationships} between inputs and outputs under controlled transformations; however, such reasoning is often informal, incomplete, and difficult to systematically capture~\cite{worstar, badimpchi}. Recent work exploring model-based testing for virtual reality programs demonstrates enhanced capabilities for testing coverage, yet limitations for detecting real-world bugs~\cite{zhu2025vrexplorer}. This observation motivates the use of \emph{metamorphic testing}, a technique that validates relationships between program behaviors rather than comparing a single output against a fixed oracle.

\subsection{Metamorphic Testing for AR Applications}

Metamorphic testing has been widely applied across domains where reliable test oracles are difficult to obtain, including scientific computing, machine learning systems, robotics, and image processing applications~\cite{mr_app_machine, mr_selection_good, mr_machine_learning,laurent2024metamorphic}. In these contexts, MT verifies whether expected relations between multiple program executions hold under specific input transformations. Prior work has also explored applying MT to AR systems. Bose et al.~\cite{bose_ar_mt} conducted a manual empirical analysis of mobile AR applications and identified seven domain-specific MRs capturing key spatial behaviors such as object scaling, occlusion consistency, and orientation stability. This study highlights that AR pipelines involve complex interactions between spatial tracking and virtual object manipulation, making behavioral properties such as MRs particularly suitable for validating AR functionality when traditional oracle-based testing becomes impractical.

\section{Approach}\label{sec:approach}

\subsection{Context Configurations for MR Generation}

Recent work on repository-level code understanding shows that structured, hierarchical representations enable LLMs to reason more effectively about large codebases while mitigating context limitations~\cite{repomain, hoang2025codewiki, oskooei2025repository}. In object-oriented systems such as Unity-based AR applications, program behavior often depends on interactions across classes and components~\cite{mr_future, llm_empirical}. These observations motivate the use of structured repository context for MR generation.

To investigate how contextual granularity influences the quality of generated MRs, we design three configurations that expose progressively richer program information to the language model. These configurations form an ablation over context, allowing us to analyze how different levels of structural information impact the validity, specificity, and diversity of generated relations.

\paragraph{A0: Method-Only Context.}
The A0 configuration represents the minimal context condition. In this setting, the language model receives only the method signature of the target function. No information about the surrounding class or repository is provided. The goal of this configuration is to measure how well an LLM can infer potential behavioral relations using only the local function interface. While this minimal prompt reduces contextual noise, it limits the model's ability to reason about program semantics beyond the method boundary.

\paragraph{A1: Flat Repository Context.}
The A1 configuration augments the method signature with a flat set of repository artifacts retrieved from the codebase. These artifacts include the repository name and type, a list of Unity-related features used in the project, class field declarations, and a set of additional class snippets selected based on method similarity. All retrieved context is provided as a single unstructured block within the prompt. This configuration allows the model to observe broader repository signals but does not explicitly encode structural relationships between components.

\paragraph{A2: Hierarchical Repository Context.}
The A2 configuration introduces a structured representation of repository context designed to reflect the hierarchical organization of object-oriented systems, commonly used to implement AR applications~\cite{hang2023ar}. The prompt contains three layers of contextual information. At the repository level, the model receives the repository name, project type, and an LLM-generated repository summary. At the class level, the prompt includes the class name, base class information, a generated class summary, and whether the class represents a Unity component. At the method level, the model receives the target method signature together with related classes identified through a dependency graph. This hierarchical structure exposes both architectural intent and localized implementation details, enabling the model to reason about program behavior at multiple abstraction levels.


\subsection{Metamorphic Relation Generation Pipeline}

\begin{figure}[t]
\centering
\footnotesize
\setlength{\fboxsep}{4pt}
\begin{tcolorbox}[colback=white,colframe=black,title=MR Generation Prompt Templates,
                  boxrule=0.5pt,arc=1mm,left=1mm,right=1mm,top=1mm,bottom=1mm]

\textbf{A0 (Method-Only).}\\
\textbf{Input:} \texttt{\{method\_signature\}}\\
\textbf{Task:} Generate candidate MRs for the target method.\\
\textbf{Output:} name, input\_transformation, output\_relation, target\_class, target\_method, source=A0

\vspace{0.4em}
\hrule
\vspace{0.4em}

\textbf{A1 (Flat Context).}\\
\textbf{Input:} \texttt{\{method\_signature\}}, \texttt{\{class\_fields\}}, \texttt{\{unity\_features\}}, \texttt{\{retrieved\_snippets\}}\\
\textbf{Task:} Generate candidate MRs using the provided flat retrieved context.\\
\textbf{Output:} name, input\_transformation, output\_relation, target\_class, target\_method, source=A1

\vspace{0.4em}
\hrule
\vspace{0.4em}

\textbf{A2 (Hierarchical Context).}\\
\textbf{Input:} \texttt{\{repo\_summary\}}, \texttt{\{class\_summary\}}, \texttt{\{inheritance\_info\}}, \texttt{\{dependency\_relations\}}, \texttt{\{method\_signature\}}\\
\textbf{Task:} Generate candidate MRs using hierarchical repository context.\\
\textbf{Output:} name, input\_transformation, output\_relation, target\_class, target\_method, source=A2

\end{tcolorbox}
\caption{Prompt structure used for candidate MR generation under the three context configurations.}
\label{fig:mr_generation_prompts}
\end{figure}

Figure~\ref{fig:mr_generation_pipeline} presents the overall MR generation workflow. Starting from Unity-based mobile AR repositories, we extract target class--method tuples $(r,c,m)$ where $r$ denotes the repository, $c$ the class, and $m$ the method, and generate candidate MRs under three context settings: A0, A1, and A2. These settings differ only in the amount and structure of context exposed to the model.

Figure~\ref{fig:mr_generation_prompts} summarizes the prompt design used in each configuration. A0 uses only the target method signature. A1 extends this with flat retrieved artifacts such as class fields, Unity features, and code snippets. A2 further introduces hierarchical repository context, including repository summaries, class summaries, inheritance information, and dependency relations. This progression allows us to study how increasing contextual richness affects MR generation quality.

Given these prompts, the language model generates one or more candidate MRs for each target method. Each MR follows a structured schema containing fields such as \texttt{name}, \texttt{input\_transformation}, \texttt{output\_relation}, \texttt{target\_class}, \texttt{target\_method}, and \texttt{source}. The \texttt{source} field records whether the candidate originated from A0, A1, or A2.

Finally, candidates produced for the same $(r,c,m)$ tuple are aggregated into its candidate set and passed to the downstream reasoning stage (Figure~\ref{fig:mr_generation_pipeline}). This design separates MR generation from MR selection: the first stage proposes multiple candidate relations, while the next stage resolves conflicts and refines them through structured reasoning.

\subsection{Reasoning-Orchestrated Debate for MR Refinement}
\begin{figure}[t]
\includegraphics[width=0.8\columnwidth]{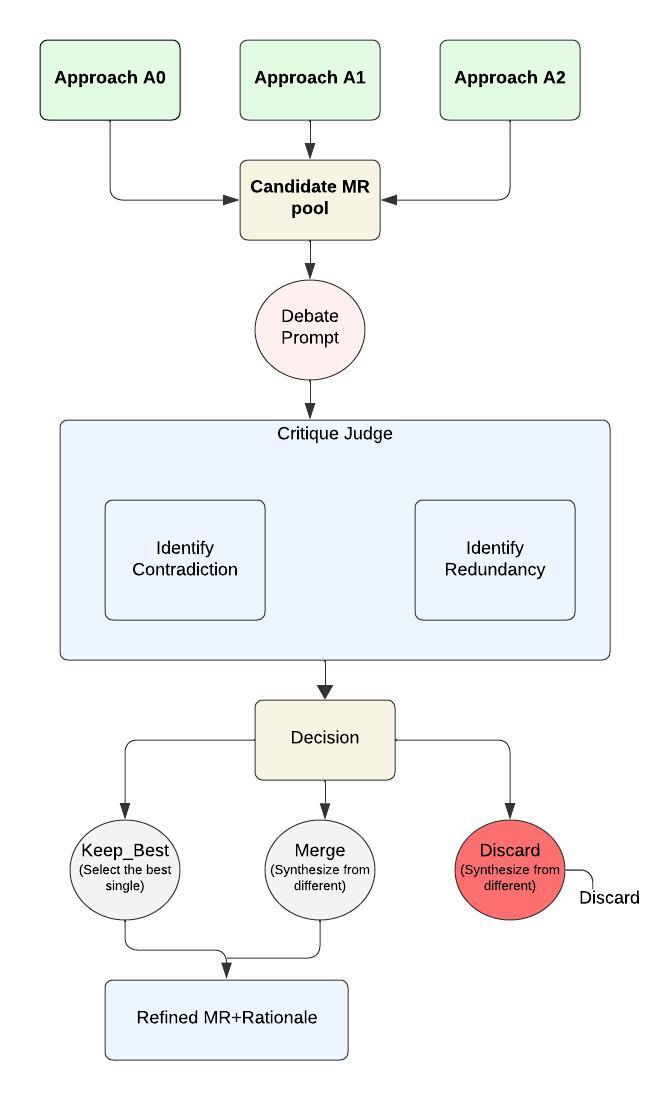}
\caption{Debate-based MR refinement workflow. Candidate MRs generated under A0, A1, and A2 are aligned by target method, assembled into a debate prompt, evaluated by a critique/judge model, and converted into a final decision: \texttt{KEEP\_BEST}, \texttt{MERGE}, or \texttt{DISCARD\_ALL}.}
\label{fig:mr_debate_pipeline}
\end{figure}

\begin{table}[h!]
\centering
\caption{Example of debate-based MR refinement (AR\_Snake\_PoseUpdate)}
\label{tab:mr-debate}
\footnotesize
\setlength{\tabcolsep}{3pt}
\renewcommand{\arraystretch}{1.15}
\begin{tabular}{|p{0.94\columnwidth}|}
\hline

\textbf{Target:} \texttt{InstantPreviewManager.UpdateLoop()} \\
\textbf{Repository:} \texttt{AR-Snake-Game-Using-Unity} \\ \hline

\textbf{A0 Assessment (Method-only):} \\
The generated MRs are overly generic and lack sufficient context about the method’s functionality. They do not capture the AR-specific behavior and are not precise enough to be meaningful or testable. \\ \hline

\textbf{A1 Assessment (Flat context):} \\
The generated MR introduces a delay-based transformation, which is a plausible scenario but does not directly align with the core functionality of the \texttt{UpdateLoop} method. It also lacks specificity in how the transformation impacts the output. \\ \hline

\textbf{A2 Assessment (Hierarchical context):} \\
The generated MR is highly specific and context-aware, correctly capturing the interaction between \texttt{InstantPreviewManager} and \texttt{InstantPreviewTrackedPoseDriver}, and how device pose updates propagate to the AR snake’s movement. \\ \hline

\textbf{Decision:} \texttt{KEEP\_BEST} \\
\textbf{Best Source:} A2 \\

\textbf{Final MR:} \\
\textbf{Name:} AR\_Snake\_PoseUpdate \\
\textbf{Input Transformation:} Apply a rotational transformation $R$ to the device pose input provided by \texttt{InstantPreviewTrackedPoseDriver}, while keeping the virtual snake anchored to the same real-world position. \\

\textbf{Output Relation:} 
(1) The snake’s world-space position must remain unchanged. \\
(2) The snake’s perceived pose (position/orientation in the camera view) must change consistently with the applied rotation $R$.

\textbf{Rationale:} The A2-generated MR is the most accurate and testable, as it reflects the core AR pipeline where real device pose updates are mapped to virtual object movement. \\

\hline
\end{tabular}
\end{table}
Figure~\ref{fig:mr_debate_pipeline} illustrates the debate-based refinement process used to resolve competing candidate MRs generated from A0, A1, and A2. For each target $(r,c,m)$, all candidate MRs are aggregated into a shared pool and passed to a critique-based reasoning stage. Table~\ref{tab:mr-debate} provides a concrete example of this process.

\paragraph{Debate Input.}
For each target method, the system collects candidate MRs generated under different context configurations. Each candidate includes its input transformation, output relation, and source label (A0, A1, or A2). These candidates often differ in abstraction level, domain awareness, and coverage of program behavior.

\paragraph{Contradiction and Redundancy Analysis.}

We choose contradiction and redundancy as the primary critique dimensions because they correspond to the two fundamental outcomes of comparing multiple MR candidates for the same target: either the candidates disagree, or they overlap~\cite{mr_base,zhou2015metamorphic}. This makes them the most appropriate dimensions for evaluating the debate stage. Structural validity, completeness, and field-level specificity are already measured at the individual-MR level in RQ1; in contrast, the debate stage must judge how candidate MRs relate to one another. For that purpose, contradiction captures unresolved conflict across candidates, while redundancy captures repeated hypotheses that add no new testing value. Together, these two dimensions directly measure whether debate is serving its intended role as a reconciliation mechanism rather than a simple selector. We validate the correctness of these decisions in section ~\ref{effect_debate}.

\begin{itemize}

\item \textbf{Contradiction:} Two MRs are contradictory if they describe incompatible output relations under similar or overlapping input transformations. For example, one MR may imply invariance under a transformation while another predicts a change in behavior. Identifying such inconsistencies is essential for ensuring logical coherence in generated specifications~\cite{mr_old}.

\item \textbf{Redundancy:} Two MRs are redundant if they express semantically equivalent relations with minor syntactic variation. Redundant relations do not contribute additional testing value and can reduce the diversity and usefulness of the MR set~\cite{zhou2015metamorphic}.
\end{itemize}

\paragraph{Critique-Based Decision.}
Based on detected contradictions and redundancies, the judge produces one of three decisions:

\begin{itemize}
\item \textbf{KEEP\_BEST}: Select a single MR that best captures the target behavior.
\item \textbf{MERGE}: Combine complementary aspects of multiple MRs into a refined relation.
\item \textbf{DISCARD\_ALL}: Remove all candidates when they are inconsistent or uninformative.
\end{itemize}

As shown in Figure~\ref{fig:mr_debate_pipeline}, this process acts as a structured filtering step that eliminates conflicting or duplicate relations before producing a final MR.

\paragraph{Refined Output.}
The output of the debate stage is a refined MR accompanied by a rationale and source attribution. This refined relation is typically more context-aware and less redundant than individual candidates. Table~\ref{tab:mr-debate} demonstrates how the debate process selects a context-aware MR (A2) over more generic alternatives, resulting in a precise and testable AR-specific relation.

Overall, the debate mechanism serves as a conflict-resolution layer that enforces consistency and reduces redundancy among candidate MRs. By explicitly reasoning over disagreements between context configurations, it produces more coherent and operationalizable relations for downstream testing.

\section{Evaluation Setup}

This section describes the dataset, evaluation metrics, and analysis procedures used to investigate the research questions introduced in Section~\ref{sec:introduction}. Our evaluation focuses on mixed methods approaches to understand how repository-level contextual information influences the structural quality and diversity of generated MRs (RQ1), how reasoning orchestration through debate refines competing MR candidates (RQ2), and the validity and usefulness of generated MRs for AR applications (RQ3).

\subsection{Dataset}

We use an existing dataset of open-source, executable mobile AR repositories introduced in prior work~\cite{BOSE_LLM_IDENTIFICATION}. The dataset consists of Unity-based mobile AR repositories written in C\# and was collected from GitHub using AR-related keywords such as \textit{``AR Foundation''}, \textit{``ARCore''}, \textit{``ARKit''}, and \textit{``Augmented Reality''}.

From repositories in this dataset, we extract class-method pairs representing potential behavioral units for analysis. In total, the dataset contains 142 repositories and 5,167 unique class--method targets. For each target method, we generate MR candidates using the three context configurations described in Section~\ref{sec:approach}. Across all configurations, the generation process produced 14,916 candidate MRs, which form the basis for our experimental analysis.

\subsection{LLMs Configuration}

We use two open-weight instruction-tuned LLMs, as generation and deliberation impose different requirements. MR generation must scale across many repository-method instances, favoring an efficient model, whereas deliberation requires stronger reasoning to compare alternatives, resolve contradictions, and refine candidate MRs. We discuss the implications of this capability asymmetry in Section~\ref{sec:discussion}.

For MR generation, we use \texttt{Llama-3.2-3B-Instruct}, selected for efficient large-scale inference and strong instruction-following performance~\cite{metallama_use, llama_review}. We set \texttt{temperature}=0.3 and \texttt{max\_new\_tokens}=1024.

For refinement, we use \texttt{Qwen2.5-Coder-32B-Instruct}, leveraging its strong code reasoning ability to resolve contradictions, remove redundancies, and refine candidate MRs~\cite{hui2024qwen2,yang2024execrepobench}. We set \texttt{temperature}=0.3 and \texttt{max\_new\_tokens}=1024.

\subsection{Structural Validity and Redundancy Metrics (RQ1)}
To evaluate the quality of generated MRs, we adopt a set of complementary metrics capturing structural validity, specificity, diversity, and coverage. We design the structural validity metrics based on minimal criteria required for a usable MR, ensuring that each relation is well-formed, sufficiently descriptive, and actionable for defining input–output transformations~\cite{mr_base,mr_coverage_new, test_addequate}. In contrast, the remaining metrics are adapted from prior work on test generation and specification mining, where coverage, specificity, and diversity are commonly used to assess the usefulness of generated artifacts~\cite {hoang2025codewiki, repomain, oskooei2025repository}.

\paragraph{Data-Driven Specificity Thresholds}

To avoid arbitrary heuristics when measuring textual specificity, we derive non-triviality thresholds from the pooled set of all generated MRs across A0, A1, and A2 ($N = 14{,}916$). For both the \texttt{input\_transformation} and \texttt{output\_relation} fields, we compute the 10th percentile (p10) character length using linear interpolation over the sorted distribution.

For a textual field $f$, non-triviality is defined as:

\[
\text{NonTrivial}(f) =
\begin{cases}
1 & \text{if } |f| \geq \text{p10} \\
0 & \text{otherwise}
\end{cases}
\]

The resulting thresholds are 30 characters for \texttt{input\_transformation} and 34 characters for \texttt{output\_relation}. These thresholds are computed once and applied consistently across all approaches.

\paragraph{Structural Validity Criteria.}

We operationalize six binary criteria to assess whether a generated MR is structurally valid and usable for specifying testable input–output transformations. In particular, field completeness (C1), target validity (C4), and source traceability (C5) ensure that each MR can be mapped to a specific code location and generation setting. Non-triviality checks (C2--C3) enforce that both the input transformation and expected output relation contain sufficient detail to be operationalized as test assertions. Finally, the non-self-referential rationale constraint (C6) prevents degenerate cases where the model simply repeats the transformation or relation without meaningful justification. 

Formally, each MR is evaluated against the following six binary criteria, where $C_i \in \{0,1\}$:

\begin{enumerate}
\item \textbf{Field Completeness (C1):} All required fields (\texttt{id}, \texttt{name}, \texttt{input\_transformation}, \texttt{output\_relation}, \texttt{target\_class}, \texttt{target\_method}, \texttt{source}) must be present and non-empty.
\item \textbf{Input Non-Triviality (C2):} The \texttt{input\_transformation} field satisfies the p10 threshold ($\geq 30$ characters).
\item \textbf{Output Non-Triviality (C3):} The \texttt{output\_relation} field satisfies the p10 threshold ($\geq 34$ characters).
\item \textbf{Target Symbol Validity (C4):} Both \texttt{target\_class} and \texttt{target\_method} must be explicitly defined.
\item \textbf{Source Traceability (C5):} The \texttt{source} field must indicate the origin configuration (A0, A1, or A2).
\item \textbf{Non-Self-Referential Rationale (C6):} The \texttt{rationale} field must not duplicate the transformation or relation text.
\end{enumerate}

\paragraph{Composite Structural Validity Score.}

To summarize structural quality, we compute a composite Structural Validity Score (SVS) as the average of the six criteria. The score ranges from 0 to 1.0, where 1.0 indicates that all structural constraints are satisfied. 
\[
\text{SVS}(mr) = \frac{1}{6} \sum_{i=1}^{6} C_i
\]\
We assign equal weight to each criterion to avoid introducing subjective bias about their relative importance. Since all criteria represent necessary (but not individually sufficient) conditions for a valid MR, equal weighting provides a neutral and interpretable aggregation that captures overall structural soundness without overfitting to any single aspect.

\paragraph{Redundancy Calculation}

We measure redundancy using exact and near-duplicate detection based on normalized MR signatures.

Exact duplicates are defined as MRs that have identical \texttt{input\allowbreak\_transformation} and \texttt{output\_relation} after normalization. We group MRs by this pair and count duplicates by keeping one instance per group and treating the remaining instances as redundant. The exact duplicate rate is computed as the proportion of such redundant instances over the total number of MRs.

Near duplicates are defined more loosely as MRs that target the same method and use the same \texttt{input\_transformation}, but may differ in their \texttt{output\_relation}. We group MRs using this relaxed signature and compute the near-duplicate rate using the same counting procedure.

\paragraph{Coverage Metrics.}

We measure coverage to quantify how broadly each approach explores the program space when generating MRs. In the context of MT, broader coverage increases the likelihood of identifying diverse and non-overlapping behavioral relations~\cite{test_addequate, mr_coverage_new}.

\begin{itemize}
\item \textbf{Total MRs Generated:} The total number of MRs reflects the overall generation capacity of an approach and its ability to propose candidate behavioral relations.

\item \textbf{Repository Coverage:} The number of unique repositories indicates how well the approach generalizes across different projects and codebases.

\item \textbf{Class Coverage:} The number of distinct classes captures how broadly different components within a system are explored.

\item \textbf{Class--Method Coverage:} The number of unique class–method pairs provides a fine-grained view of coverage, indicating how many specific program locations are associated with generated MRs. To avoid ambiguity, methods are identified using \texttt{ClassName.methodName}.
\end{itemize}

\paragraph{Unique MR Rate.}

Finally, we compute the unique MR rate as:
$\mathrm{UniqueMRRate}(A)=1-\mathrm{ExactDupRate}(A)$.

This metric reflects the proportion of structurally distinct MRs generated by each configuration.

\subsection{Debate Outcome Analysis (RQ2)}

To analyze how reasoning orchestration refines competing MR candidates, we examine the outcomes of the debate stage introduced in Section~\ref{sec:approach}. We record the distribution of these decisions, identify which context configuration most frequently produces the winning MR, and analyze conflict signals such as contradictions or redundancies detected during debate. These measurements allow us to quantify how reasoning orchestration consolidates candidate relations into a smaller set of refined MRs.

To validate the correctness of these judge decisions, two authors independently rated a stratified random sample of 500 debate outcomes for KEEP\_BEST ($n = 205$), MERGE ($n = 213$), and DISCARD\_ALL ($n = 82$). The raters scored each decision as Correct (2), Debatable (1), or Incorrect (0) given the judge's assessments and detected conflict signals. We calculate Cohen's $\kappa$ over the 500 MR decisions to measure the agreement between raters.%

\subsection{Test Oracle Study (RQ3)}
\begin{table*}[h!]
\centering
\caption{Evaluation Criteria for Test Oracle Study}
\small
\label{tab:rq3_scale}
\begin{tabular}{p{2.5cm} p{5cm} p{7cm}}
\toprule
\textbf{Dimension} & \textbf{Key Question} & \textbf{Evaluation Criteria} \\
\midrule

\textbf{Validity} 
& Is the MR logically correct and meaningful? 
& (1) Transformation defines a clear input change; 
(2) Output relation correctly relates source and follow-up outputs; 
(3) Transformation and relation are logically consistent; 
(4) Violations indicate real faults \\

\midrule

\textbf{Testability} 
& Can the MR be directly implemented as a test? 
& (1) Transformation is programmatically applicable; 
(2) Output relation is expressible as an assertion; 
(3) Terms are unambiguous; 
(4) No interpretation is required for implementation \\

\midrule

\textbf{AR Specificity} 
& Does the MR reflect AR-specific behavior? 
& (1) Uses AR concepts (e.g., pose, alignment, occlusion); 
(2) Not trivially applicable to non-AR systems; 
(3) Targets AR-related methods or behaviors; 
(4) Reflects real-world interaction or spatial logic \\

\bottomrule
\end{tabular}
\end{table*}
To evaluate whether debate-refined relations can serve as practical test oracles, we conduct a manual evaluation study on a curated subset of MRs. Starting from the full set of debate-refined MRs ($N = 3{,}760$), the first author manually reviewed all candidates and selected 141 MRs that capture AR-specific behaviors. The selection is based on the semantic content of each MR, including the input transformation, output relation, and associated program context (target class and method).

We note that MR names are not used as a filtering criterion, as prior work has shown that surface-level identifiers are often unreliable indicators of semantic intent in generated artifacts~\cite{tufano2022using,chen2021evaluating}. Instead, selection follows explicit criteria grounded in established AR/3D interaction concepts~\cite{bowmanhci}: MRs capturing pose propagation, spatial transformations, rendering consistency, or camera- and sensor-driven behavior are included; those involving generic state changes, UI-only logic, or behavior with no AR grounding are excluded. Full criteria and sample exclusions are in our artifact.

\paragraph{Evaluation Dimensions.}
We evaluate each MR along three dimensions: \emph{validity}, \emph{testability}, and \emph{AR specificity}. These dimensions are grounded in prior work on test oracle design and specification quality, where useful oracles must be logically correct, executable, and contextually relevant~\cite{testoraclemodel,oracle_emprical}. Note that testability here denotes a property of the MR specification itself, whether its transformation and relation can be operationalized as a programmatic assertion, rather than the testability of the system under test. 

\paragraph{Scoring Scheme.}
Following standard practices in empirical software engineering for qualitative artifact evaluation, we adopt a 3-point ordinal scale (0--2) for each dimension~\cite{artifactse}. Such coarse-grained scales are commonly used to balance annotation consistency and discriminative power in human evaluation studies~\cite{rojas2017whole,jiang2007deckard}. The scale definitions are operationalized using structured checklists, as summarized in Table~\ref{tab:rq3_scale}.

\paragraph{Annotation Procedure.}
Three independent raters (Rater 1, Rater 2, and Rater 3) evaluated all 141 MRs using the predefined criteria. Each rater independently assigned scores across all dimensions without coordination. This process follows standard multi-rater evaluation protocols used in empirical software engineering studies.

\paragraph{Inter-Rater Agreement.}
To assess annotation reliability, we compute Fleiss’ $\kappa$ for multi-rater agreement and Krippendorff’s $\alpha$ (ordinal) to account for disagreement magnitude. In addition, we report exact agreement and adjacent agreement (difference $\leq 1$), which is appropriate for ordinal scales with skewed distributions. Pairwise weighted $\kappa$ is also computed to analyze agreement between individual raters.

\section{Result}

\subsection{RQ1: Effect of Context Granularity}
\begin{table}[h!]
\centering
\caption{Effect of Context Granularity on Specificity and Diversity}
\label{tab:rq1_specificity_diversity}
\begin{tabular}{lccc}
\toprule
\textbf{Metric} & \textbf{A0} & \textbf{A1} & \textbf{A2} \\
\midrule
\multicolumn{4}{c}{\textit{Input Transformation Specificity}} \\
Mean Length & 60.8 & 72.6 & 75.0 \\
Median Length & 47 & 63 & 69 \\
\midrule
\multicolumn{4}{c}{\textit{Output Relation Specificity}} \\
Mean Length & 68.0 & 77.5 & 73.5 \\
Median Length & 57 & 72 & 70 \\
\midrule
\multicolumn{4}{c}{\textit{Diversity}} \\
Exact Duplicate Rate & 18.9\% & 14.8\% & 13.3\% \\
Near Duplicate Rate & 10.8\% & 7.2\% & 6.7\% \\
Unique MR Rate & 81.1\% & 85.2\% & 86.7\% \\
\bottomrule
\end{tabular}
\end{table}

\begin{table*}[h!]
\centering
\caption{Effect of Context Granularity on Structural Validity and Coverage}
\label{tab:rq1_structural}
\begin{tabular}{lccc}
\toprule
\textbf{Metric} & \textbf{A0 (Method-Only)} & \textbf{A1 (Method+Class)} & \textbf{A2 (Hierarchical)} \\
\midrule
\multicolumn{4}{c}{\textit{Coverage}} \\
Total MRs Generated & 4,046 & 3,866 & 7,004 \\
Unique Classes Covered & 326 & 305 & 474 \\
Unique Class.Method Pairs & 3,144 & 2,933 & 5,167 \\
\midrule
\multicolumn{4}{c}{\textit{Overall Structural Validity}} \\
Mean SVS & 0.9624 & 0.9615 & 0.9702 \\
\% SVS = 1.0 & 82.45\% & 86.26\% & 85.21\% \\
\midrule
\multicolumn{4}{c}{\textit{Selected Per-Criterion Pass Rates}} \\
Input Non-Triviality(\% $\geq$ p10 Threshold) & 86.7\% & 90.4\% & 92.43\% \\
Output Non-Triviality(\% $\geq$ p10 Threshold) & 90.78\% & 91.72\% & 92.00\% \\
Target Validity & 100\% & 100\% & 100\% \\
Source Traceability & 100\% & 96.25\% & 100\% \\
\bottomrule
\end{tabular}
\end{table*}
Table~\ref{tab:rq1_specificity_diversity} and 
Table~\ref{tab:rq1_structural} show that hierarchical context (A2) achieves the strongest coverage, specificity, and diversity overall, 
Though the relationship between context volume and quality is not uniformly monotonic across all criteria.

\paragraph{Coverage.}
A2 achieves the highest coverage across all dimensions, including unique classes (474), and class--method pairs (5,167), compared to A0 and A1 (Table~\ref{tab:rq1_structural}). This indicates that hierarchical context enables the model to generalize across a broader set of program elements, beyond simply generating more MRs.

\paragraph{Structural Validity.}
All configurations maintain high structural validity (SVS $\approx$ 0.96--0.97), but the relationship between context and quality is not monotonic. A1, despite providing more raw context than A0, degrades source traceability to 96.25\%---the 
only configuration falling below 100\% on this criterion. This suggests unstructured flat retrieval introduces noise that disrupts schema adherence without adding behavioral signal. Hierarchical context (A2) recovers this and goes further, achieving the highest non-triviality rates for both input (92.43\%) and output relations (92.00\%), indicating that structured context enables more meaningful transformations rather than merely longer outputs.

\paragraph{Specificity and Diversity.}
As shown in Table~\ref{tab:rq1_specificity_diversity}, A2 improves input specificity (highest mean and median lengths, and 92.4\% above the threshold) while maintaining competitive output specificity. In addition, A2 reduces redundancy (lowest exact and near-duplicate rates) and achieves the highest unique MR rate (86.7\%), indicating more diverse and distinct MRs.

\begin{tcolorbox}[colframe=blue!50!black, colback=blue!10!white,title=RQ1 Summary]
Hierarchical context (A2) achieves the highest coverage (5,167 class--method pairs) and diversity
(86.7\% unique MRs) while maintaining high structural validity
(SVS~$\approx$~0.96). We also observed flat retrieval (A1) degrades source traceability to 96.25\%---below both A0 and A2---showing that context \emph{organization} matters as much as context \emph{volume}.
\end{tcolorbox}

\subsection{RQ2: Effect of Multi-Agent Debate}\label{effect_debate}
\begin{table}[h!]
\centering
\caption{RQ2: Conflict Identification and Resolution in Multi-Agent Debate ($N=4496$)}
\label{tab:rq2_conflict_resolution}
\begin{tabular}{lrr}
\toprule
\textbf{Metric} & \textbf{Count} & \textbf{Rate} \\
\midrule
Contradictions detected & 3550 & 79.0\% \\
Redundancies detected & 2301 & 51.2\% \\
No issues detected & 442 & 9.8\% \\
\bottomrule
\end{tabular}
\end{table}
\begin{table}[t]
\centering
\caption{RQ2: Winning Source Distribution After Debate ($N=3760$ ($4496-736$ DISCARD\_ALL))}
\label{tab:rq2_winner}
\begin{tabular}{lrr}
\toprule
\textbf{Source} & \textbf{Count} & \textbf{Rate} \\
\midrule
A2 (Hierarchical) & 3316 & 88.2\% \\
MERGED & 388 & 10.3\% \\
A1 & 47 & 1.2\% \\
A0 & 9 & 0.2\% \\
\bottomrule
\end{tabular}
\end{table}
Table~\ref{tab:rq2_conflict_resolution}, Table~\ref{tab:rq2_winner}, and Table~\ref{tab:rq2_impact} summarize the role of multi-agent debate in identifying conflicts and improving MR quality.

\paragraph{Conflict Identification and Resolution.} Deliberation was applied only to the $4,496$ targets whose candidate MRs originated from at least two context configurations, since debate requires competing candidates; single-source targets were excluded. As shown in Table~\ref{tab:rq2_conflict_resolution}, conflicts are prevalent among candidate MRs, with contradictions detected in 79.0\% and redundancies in 51.2\% of cases, while only 9.8\% exhibit no issues. This highlights that independently generated MRs frequently disagree or overlap.

\paragraph{Preference for Context-Rich MRs.}
Table~\ref{tab:rq2_winner} shows that the debate process overwhelmingly favors MRs generated with hierarchical context. A2 accounts for 88.2\% of final selections, while A1 and A0 contribute minimally (1.2\% and 0.2\%, respectively). Additionally, 10.3\% of cases result in merged outputs, suggesting that combining complementary information across sources can further improve MR quality. These results demonstrate that richer contextual information consistently leads to more reliable and preferred MRs.

\paragraph{Impact on MR Quality.}
The end-to-end impact of debate is shown in Table~\ref{tab:rq2_impact}. Debate-refined MRs are more detailed, with higher average and median lengths for both input transformations (89.3 / 83.0) and output relations (101.2 / 95.0) compared to all individual configurations. At the same time, redundancy is substantially reduced, with the exact duplicate rate dropping from 13.3\% (A2) to 1.3\%. Although the total number of MRs decreases to 3,760 (the 4,496 debate groups minus 736 DISCARD\_ALL decisions), this reflects effective filtering and consolidation rather than loss of useful information.

\paragraph{Manual Validation of Judge Decisions}
The two raters agreed on 94.2\% of decisions ($\kappa$ = 0.865; weighted $\kappa$ = 0.87), with all disagreements adjacent on the scale (Correct vs. Debatable). Only 0.6\% ($n = 3$) of decisions were flagged as Incorrect by either rater; the remainder split between clearly correct (69\%) and debatable (31\%). KEEP\_BEST (94–98\%) and DISCARD\_ALL (89–94\%) were overwhelmingly rated clearly correct, indicating the judge is trustworthy for comparative and dismissive decisions. MERGE is the outlier: only 31–37\% of MERGE decisions were rated unambiguously correct. The majority of these were due to the LLM judge making the final decision to MERGE when KEEP\_BEST would have been the simpler outcome.

\begin{tcolorbox}[colframe=blue!50!black, colback=blue!10!white,title=RQ2 Summary]
Multi-agent debate resolves conflicts in 90\% of cases and overwhelmingly selects A2-derived MRs (88.2\%), reducing duplicate rates from 13.3\% to 1.3\% while producing more detailed and refined relations.
\end{tcolorbox}

\begin{table*}[h!]
\centering
\small
\caption{End-to-End Impact of Debate on MR Quality}
\label{tab:rq2_impact}
\begin{tabular}{lcccc}
\toprule
\textbf{Metric} & \textbf{A0} & \textbf{A1} & \textbf{A2} & \textbf{Debate-Refined} \\
\midrule
Total MRs & 4046 & 3866 & 7004 & 3760 \\
Avg Transformation Length & 60.8 & 72.6 & 75.0 & \textbf{89.3} \\
Median Transformation Length & 47.0 & 63.0 & 69.0 & \textbf{83.0} \\
Avg Relation Length & 68.0 & 77.5 & 73.5 & \textbf{101.2} \\
Median Relation Length & 57.0 & 72.0 & 70.0 & \textbf{95.0} \\
Exact Duplicate Rate & 18.9\% & 14.8\% & 13.3\% & \textbf{1.3\%} \\
\bottomrule
\end{tabular}
\end{table*}

\subsection{RQ3: Generated MR Practicality}

\begin{table}[t]
\centering
\small
\caption{Quality of Debate-Refined MRs}
\label{tab:rq3_quality}
\begin{tabular}{lcccc}
\toprule
\textbf{Dimension} & \textbf{Mean} & \textbf{SD} & \textbf{Majority High} & \textbf{Consensus High} \\
\midrule
Validity & 1.80 & 0.35 & 86.5\% & 67.4\% \\
AR Specificity & 1.68 & 0.45 & 82.9\% & 48.2\% \\
Testability & 1.82 & 0.29 & 88.0\% & 63.8\% \\
\midrule
Composite & 1.77 & 0.23 & -- & -- \\
\bottomrule
\end{tabular}
\end{table}



To examine the suitability of generated MRs for testing real-world AR programs, we conducted a manual oracle study.

\paragraph{Inter-Rater Reliability.}
Three raters independently evaluated 141 debate-refined MRs on validity, AR specificity, and testability. Overall agreement is in the slight-to-fair range, with Fleiss' $\kappa$ values of 0.33 for validity, 0.22 for AR specificity, and 0.19 for testability. However, these values should be interpreted together with the high adjacent agreement rates of 94.3\%, 95.7\%, and 96.5\%, respectively, indicating that disagreements were usually limited to a one-level difference on the 0--2 scale. Pairwise weighted agreement reveals a clear structure: Rater~1 and Rater~2 are strongly aligned across all three dimensions ($\kappa_w=0.73$, $0.73$, and $0.58$), whereas agreement involving Rater~3 is much lower, especially for testability ($\kappa_w=-0.01$ between Rater~1 and Rater~3). This pattern suggests that the lower multi-rater coefficients are driven less by random disagreement and more by a systematic difference in rating strictness, particularly for AR specificity and testability.

\paragraph{Quantitative Findings.}
Table~\ref{tab:rq3_quality} shows that validity and testability are the strongest dimensions, with mean scores of 1.80 and 1.82, and majority-high rates of 86.5\% and 88.0\%, respectively. This indicates that most debate-refined MRs are both logically sound and sufficiently well-specified for human raters to judge them as directly implementable as test conditions without requiring additional interpretation. AR specificity is lower and more variable (mean = 1.68, SD = 0.45), which is consistent with the fact that the curated set spans both AR-core behaviors and application-layer logic. Even so, 82.9\% of MRs receive majority-high AR-specificity scores. Across all three dimensions, 87 of 141 MRs (61.7\%) satisfy the high-quality criterion, and 35 (24.8\%) achieve top-tier quality across all dimensions.

\paragraph{Effect of Deliberation Strategy.}
The debate strategy also affects oracle quality. MRs produced through \textsc{merge} outperform \textsc{keep-best} on every dimension, with a higher composite score (1.81 vs.\ 1.73), higher validity (1.85 vs.\ 1.77), higher AR specificity (1.70 vs.\ 1.66), and higher testability (1.87 vs.\ 1.78). This indicates that synthesizing information from competing candidates yields stronger oracle specifications than simply retaining a single candidate.

\begin{tcolorbox}[colframe=blue!50!black, colback=blue!10!white,title=RQ3 Summary]
Debate-refined MRs achieve high validity (1.80) and testability (1.82), with 61.7\% rated high-quality across all dimensions, indicating that most generated relations are sufficiently valid 
and well-specified to serve as practical test oracle specifications.
\end{tcolorbox}

\section{Discussion}\label{sec:discussion}

Based on our findings, we provide implications for LLM MR generation, testing AR applications, and metamorphic testing in practice.

\subsection{Implications for LLM-Based MR Generation}


Our results surface implications for LLM-based MR generation that go beyond the intuition that ``more context is better''.

\textbf{Context Configurations.} We observed flat retrieval (A1) actually \emph{degrades} source traceability below both the method-only
baseline (A0) and the hierarchical configuration (A2), dropping from 100\% to 96.25\%
(Table~\ref{tab:rq1_structural}), failing to map generated MRs to the source code. This occurs because unstructured retrieval
injects loosely related snippets that disrupt schema adherence without adding behavioral
signal. Hierarchical context (A2) recovers this and additionally improves specificity
(92.4\% non-trivial input transformations vs.\ 86.7\% for A0) and diversity
(86.7\% unique MR rate vs.\ 81.1\%). Taken together, these results suggest that
\emph{how} context is organized matters as much as \emph{how much} context is provided.

Repository-level LLM tasks more broadly ---i.e.,
code search, bug localization, specification mining --- face the same tension between
providing sufficient context and maintaining structural coherence~\cite{oskooei2025repository, repomain}. Our ablation shows hierarchical
organization, which preserves relationships between repository, class, and method layers,
outperforms flat retrieval, even when the flat approach has access to the same underlying
artifacts. LLM-based MR generation pipelines over large codebases should treat context structure as a first-class design decision, not an implementation detail.

\textbf{Debate-Based Refinement.} A second key finding is that MR generation is inherently a multi-perspective problem.
Candidates produced independently under A0, A1, and A2 exhibit contradictions in 79.0\%
of cases and redundancies in 51.2\%, confirming that no single context view captures the full behavioral surface of a target method. Single-pass generation pipelines, regardless of context richness, will therefore produce unreliable candidate sets.
Reasoning-based deliberation is not an optional refinement step; it is structurally necessary. The debate mechanism reduces exact duplicates from 13.3\% to 1.3\% and consistently selects or synthesizes context-aware relations, with merged outputs outperforming single-best selections on every quality dimension in our oracle study (composite score 1.81 vs.\ 1.73). 


\subsection{Implications for Practitioners}\label{impli_PRAC}
Despite strong overall results, 38.3\% of evaluated MRs did not meet our high-quality threshold across all three dimensions, reflecting two practical realities practitioners should anticipate.

\textbf{Human-in-the-loop.}
MRs passing structural validity checks can still be too abstract for direct instantiation, particularly for application-layer behaviors lacking clear AR-specific semantics. AR specificity is the most variable dimension (mean~1.68, SD~0.45), confirming the pipeline is most reliable for core AR behaviors like pose propagation, spatial transformations, rendering consistency and less so for higher-level logic. Practitioners should treat output as a ranked candidate set requiring lightweight human review.

\textbf{Model asymmetry and generalizability.}
We intentionally pair a smaller generation model (Llama-3.2-3B) with a larger deliberation model (Qwen2.5-Coder-32B) to balance scalability with reasoning quality. This raises the question of whether the judge's strong preference for A2 (88.2\%) partly reflects shared biases with hierarchical prompting. The MERGE outcomes (10.3\%), whose merged MRs consistently outperform KEEP\_BEST despite diluting the A2 signal, and the independent
human oracle study both provide counterevidence. Future work should evaluate deliberation with matched-capability models to isolate context structure from model capacity. Although our pipeline is domain-agnostic in design, all 142 repositories are Unity-based C\# projects; extension to domains such as robotics or embedded systems remains a necessary validation step.

\subsection{Execution-Based Validation}
\label{subsec:execution_validation}

To examine whether generated MRs can be operationalized as executable
tests and whether they detect injected faults, we conducted a preliminary
case study focused on execution-based validation. 

\paragraph{Method:} We randomly selected five
debate-refined MRs from different AR repositories. The first author translated each natural
language relation into property-based assertions over independently checkable
properties, reviewed by a second author. For each MR, we executed 10,000 randomly generated test cases on the
correct implementation and on mutated versions of the target method. Following prior work that uses mutation analysis to evaluate generated MRs across
domains~\cite{just2009evaluating,murphy2011effective,xiao2022metamorphic}, we
applied four commonly used mutation operators: Arithmetic Operator
Replacement (AOR), Unary Operator Insertion (UOI), Scalar Variable Replacement
(SVR), and Statement Deletion (SDL). These mutations were applied mechanically,
without using knowledge of the MR properties. Across the five randomly selected MRs, we applied 30 mutants
in total.

Figure~\ref{fig:mr_to_pbt_example} illustrates one such translation for
\texttt{Frame.\_Draw2DBody()}. The generated MR perturbs GL-space coordinates
and expects the resulting world-space position to change consistently under a
fixed camera configuration and depth. Rather than treating the MR only as a
semantic relation, we operationalize it as a reusable property-based test over
transformed inputs and relational output assertions.

\begin{figure}[t]
\centering
\begin{tcolorbox}[colback=white,colframe=black,
                  title=Illustrative MR-to-PBT Translation,
                  boxrule=0.5pt,arc=1mm,left=1mm,right=1mm,top=1mm,bottom=1mm]
\footnotesize
\textbf{Target Method:} \texttt{Frame.\_Draw2DBody()} \\[0.3em]
\textbf{Relevant Code Fragment:}
\begin{verbatim}
Vector3 glCoord = pair.Value.Coordinate2D;
Vector3 worldCoord = new Vector3((glCoord.x + 1) / 2,
    (glCoord.y + 1) / 2, 3);
m_skeletonPointObject[(int)pair.Key].transform.position =
    m_skeletonCamera.ViewportToWorldPoint(worldCoord);
\end{verbatim}

\textbf{Generated MR:} \\[0.2em]
\textit{Input Transformation:} Perturb the input GL coordinates $(x, y)$ by a
bounded non-zero offset $(\delta x, \delta y)$ such that both the original and
perturbed coordinates remain within the valid range $[-1,1]^2$. \\[0.3em]
\textit{Output Relation:} For a fixed camera configuration and fixed depth
$z=3$, perturbing the GL-space input should produce a corresponding displacement
in the world-space position returned by \texttt{ViewportToWorldPoint}. The
displacement should be directionally consistent with the perturbation, and
repeated equal perturbation steps should yield approximately equal changes in
world-space position. \\[0.3em]
\textbf{Derived PBT Property:}
\begin{verbatim}
For all (x,y) in [-1,1]^2 and (dx,dy) such that
(x+dx, y+dy) and (x+2dx, y+2dy) remain in [-1,1]^2:

  p   = ViewportToWorld((x+1)/2,      (y+1)/2,      3)
  p'  = ViewportToWorld((x+dx+1)/2,   (y+dy+1)/2,   3)
  p'' = ViewportToWorld((x+2dx+1)/2,  (y+2dy+1)/2,  3)

  assert ||p' - p|| > 0 when (dx,dy) != (0,0)
  assert ||(p'' - p') - (p' - p)|| < epsilon
\end{verbatim}
\end{tcolorbox}
\caption{Example of translating a generated MR into a property-based testing
specification.}
\label{fig:mr_to_pbt_example}
\end{figure}

\begin{table*}[t]
\centering
\caption{Translation of five generated MRs into executable properties and
mutation checks.}
\label{tab:mr_property_mutation_all}
\small
\renewcommand{\arraystretch}{1.12}
\begin{tabularx}{\textwidth}{
    >{\raggedright\arraybackslash}p{0.18\textwidth}
    >{\raggedright\arraybackslash}X
    >{\raggedright\arraybackslash}p{0.30\textwidth}
}
\toprule
\textbf{MR Type} & \textbf{Executable Properties} & \textbf{Detected Mutations / Fault Types} \\
\midrule

GL-coordinate perturbation
&
Non-zero GL perturbation should produce non-zero world-space displacement;
equal GL perturbation steps should produce equal world-space steps;
world-space displacement direction should match GL-space perturbation;
single-axis perturbation should not affect the orthogonal axis.
&
Killed mutations that negated \texttt{gl\_x}, negated \texttt{gl\_y}, or
replaced \texttt{gl\_x} with \texttt{gl\_y}; detects wrong-sign arithmetic
and axis-confusion faults.
\\

\midrule

Position reset
&
Moving an object by $+v$ and then $-v$ should return it to the original
position; zero displacement should preserve position; sequential moves should
compose additively; displacement magnitude should be preserved.
&
Killed mutations that scaled one displacement component, substituted one axis
for another, or removed the update on one axis; detects scaling, variable
substitution, and missing-update faults.
\\

\midrule

Rotation magnitude scaling
&
Scaling the shake amount by factor $k$ should scale the resulting rotation by
$k$; zero amount should produce zero rotation; larger input magnitude should
produce larger rotation magnitude; axis-specific input should affect only the
corresponding rotation axis.
&
Killed arithmetic mutations replacing multiplicative scaling with additive
updates; detects violations of proportionality in rotation response.
\\

\midrule

Camera position transformation
&
Moving the camera by displacement $d$ should shift each trackable's
camera-relative position by $-d$; zero camera movement should preserve relative
position; sequential translations should compose additively; trackable
world-space position and inter-trackable distances should remain invariant.
&
Killed mutations changing subtraction to addition, negating the relative
position, dropping a camera-coordinate update, returning camera position instead
of trackable position, or scaling the relative vector; detects incorrect
coordinate-frame transformations.
\\

\midrule

Raycast position shift
&
A small screen-coordinate shift should produce a bounded movement in the 3D hit
point; zero shift should be deterministic; hit displacement should remain
ratio-consistent across shift magnitudes; X-direction screen shifts should move
the hit point consistently in world space; doubling the shift should
approximately double the hit displacement.
&
Killed mutations that negated the ray's X direction or substituted
\texttt{screen\_y} for \texttt{screen\_x}; detects wrong-direction ray mapping
and screen-axis confusion.
\\

\bottomrule
\end{tabularx}
\end{table*}

\paragraph{Results:}  Table~\ref{tab:mr_property_mutation_all} summarizes how each MR was
converted into executable properties and what types of injected faults those
properties detected. The generated property-based tests killed 16 out of 30 mutants (53\%). The remaining 14
mutants did not violate the corresponding MR properties because they preserved
the relational behavior being tested. For example, some mutations changed
absolute offsets, centering constants, waveform details, or isometric structure,
but did not change the input-output relation specified by the MR. These cases
were therefore treated as equivalent with respect to the tested metamorphic
property. \textbf{After excluding such equivalent mutants, all non-equivalent mutants (16/16, 100\%)
were detected}.

\paragraph{Implication:} These results provide preliminary evidence that the generated MRs are not only structurally valid and human-rated as testable, but can also be translated into executable tests that detect seeded behavioral faults such as wrong-sign arithmetic, axis confusion, variable substitution errors, scaling errors, etc.

\section{Threats to Validity}

\paragraph{Internal Validity.}
Our structural metrics rely on design choices such as p10-based thresholds and equal weighting in SVS. We mitigate this by deriving thresholds from the pooled dataset and reporting both aggregate and per-criterion results. The deliberation stage additionally depends on LLM reasoning, which may introduce variability; we address this by reporting decision distributions and conflict signals across a large number of debates. 

\paragraph{Construct Validity.}
The three oracle dimensions (validity, testability, AR specificity) may not capture all aspects of usefulness in real-world development workflows. We mitigate this by grounding them in prior work~\cite{testoraclemodel, oracle_emprical} and using structured scoring checklists (Table~\ref{tab:rq3_scale}) to ensure consistent interpretation.

\paragraph{External Validity.}
Results are based on 142 Unity C\# repositories and two LLMs; generalizability to other domains beyond AR and models is limited. We mitigate this by designing the pipeline to be domain- and LLM-agnostic. Our preliminary case study is also limited to five repositories, which future work will scale.

\paragraph{Conclusion Validity.}
The oracle study involves three raters and an ordinal scale, which may introduce subjectivity. We mitigate this with multiple agreement metrics (Fleiss'~$\kappa$, Krippendorff's~$\alpha$, exact and adjacent agreement) and by analyzing score distributions to account for skewed ratings.

\section{Future Work}
This work opens several future research directions. First, generated MRs can be automatically translated into executable test cases (e.g., property-based or unit tests), enabling end-to-end integration into testing pipelines. Incorporating runtime feedback (e.g., test failures or execution traces) could further refine and validate generated relations. Extending to other domains that often leverage metamorphic testing beyond AR (e.g., robotics~\cite{laurent2024metamorphic} or simulation systems~\cite{ahlgren2021testing}) and incorporating other models in our evaluation pipeline (e.g., GPT-4 or Claude) would help assess the generalizability of our approach. Finally, improving domain grounding—such as leveraging API semantics or engine-specific knowledge—may further enhance the precision of generated MRs for application-layer behaviors.

\section{Conclusion}

We present a repository-aware pipeline for generating and refining MRs through structured context and reasoning-based deliberation. We evaluate our approach on 142 real-world AR codebases. Our results show hierarchical repository context improves coverage and diversity, while deliberation effectively resolves conflicts and reduces redundancy. Human inspection further demonstrates refined MRs are largely valid, testable, and usable as test oracles, and a preliminary case study ($n = 5$) shows generated MRs can be translated into property-based tests that surface bugs in real-world AR code. These findings highlight the potential of combining repository-level context with reasoning orchestration to enable scalable and reliable oracle generation for complex software systems.








\section{Data Availability}

All data and study artifacts, including our evaluation pipeline scripts, prompt templates, etc., are publicly available to support reproducibility and reuse.\footnote{\url{https://github.com/brintodibyendu/MR_DISCOVERY_ASE}}

\bibliographystyle{ACM-Reference-Format}
\bibliography{sample-base}

\appendix

\end{document}